\documentclass[a4paper,fleqn,usenatbib]{mnras}

\newcommand{\gppr}{\stackrel{>}{\scriptstyle \sim}}
\newcommand{\gappr}{\raisebox{-0.4ex}{$\gppr$}}

\newcommand{\Lsun}{L_{\odot}}
\newcommand{\Msun}{M_{\odot}}
\newcommand{\Rsun}{R_{\odot}}
\newcommand{\Teff}{T_{eff}}
\usepackage[T1]{fontenc}
\usepackage{ae,aecompl}

\usepackage{graphicx}	
\usepackage{amsmath}	
\usepackage{amssymb}	
\usepackage{threeparttable} 
\usepackage{placeins}
\usepackage{color}

\title[Accretion in SR\,12\,C]{Accretion signatures in the X-shooter spectrum of
  the substellar companion to SR12}
\author[A. Santamar\'ia-Miranda et al.]{
Alejandro Santamar\'ia-Miranda,$^{1,2,6,7}$\thanks{E-mail: asghaloth@gmail.com}
Claudio C\'aceres,$^{3,6}$ Matthias R. Schreiber,$^{1,6,7}$ 
\newauthor Adam Hardy,$^{1,7}$ Amelia Bayo,$^{1,6}$ Steven G. Parsons,$^{4}$ Mariusz Gromadzki$^{5}$ \newauthor Aurora Bel\'en Aguayo Villegas$^{1,6}$
\\
$^{1}$Instituto de F\'isica y Astronom\'ia, Universidad de Valpara\'iso, Av. Gran Breta\~na 1111, 5030 Casilla, Valpara\'iso, Chile\\
$^{2}$European Southern Observatory, Ave. Alonso de Cordova 3107, Casilla 19, 19001, Santiago, Chile \\
$^{3}$Departamento de Ciencias Fisicas, Facultad de Ciencias exactas, Universidad Andres Bello. \\Av. Fernandez Concha 700, Las Condes, Santiago de Chile\\
$^{4}$Department of Physics and Astronomy, University of Sheffield, Sheffield, 53 7RH, UK\\
$^{5}$Warsaw University Astronomical Observatory, Al. Ujazdowskie 4, 00-478 Warszawa, Poland\\ 
$^{6}$N\'ucleo Milenio Formaci\'on Planetaria - NPF, Universidad de Valpara\'iso, Av. Gran Breta\~na 1111,  Valpara\'iso, Chile \\
$^{7}$Millennium Nucleus "Protoplanetary Disks in ALMA Early Science", Universidad de Valpara\'iso, Chile\\
}

\date{Accepted 2017 December 22. Received 2017 December 19; in original form 2017 May 29}

\pubyear{2017}

\begin{document}
\label{firstpage}
\pagerange{\pageref{firstpage}--\pageref{lastpage}}
\maketitle

\begin{abstract}
{About a dozen substellar companions orbiting young 
stellar objects or pre-main sequence stars at several hundred au have been identified in the last decade. 
These objects are interesting both due to the uncertainties surrounding their formation, and
because their large separation from the host star 
offers the potential to study the atmospheres of
young giant planets and brown dwarfs. 
Here, we present X-shooter spectroscopy of SR\,12\,C, 
a $\sim\,2\,$Myrs young brown dwarf orbiting SR\,12 at an orbital separation of
1083 au. We determine the spectral type, gravity, and effective temperature via comparison with models and observational templates of young brown dwarfs. In addition, we detect and characterize accretion using several accretion tracers. 
We find SR\,12\,C to be a brown dwarf of spectral type L0 $\pm$ 1, $\log$ g
= 4 $\pm$ 0.5, an effective temperature of 2600 $\pm$ 100 {K}. 
Our spectra provide clear evidence for accretion at a rate of 
$\sim\,10^{-10}\Msun/\mathrm{yr}$.
This makes SR\,12 one of the few sub-stellar companions with a reliable 
estimate for its accretion rate. A comparison of the ages and accretion rates of sub-stellar
companions with young isolated brown dwarfs does not reveal any significant
differences. If further accretion rate measurements of a large number of
substellar companions can confirm this trend, this would hint towards a similar formation mechanism 
for substellar companions at large separations and isolated brown dwarfs.
} 
\end{abstract}

\begin{keywords}
brown dwarfs -- stars: pre-main-sequence -- accretion, accretion discs-\end{keywords}



\section{Introduction}

Planetary mass or brown dwarf companions which 
orbit their host stars at separations exceeding 100\,au are 
intriguing objects that may provide the potential to perform detailed
investigations of the atmospheres of young giant planets and brown dwarfs.  
However, how these interesting sub-stellar companions (SSCs) formed remains a mystery. 
At separations of several hundred au (i.e. much larger than most known 
exoplanets or brown dwarf companions) the currently most accepted planet
formation scenario of in-situ core accretion is unlikely to
occur. This is because the time required for core
growth at these large distances significantly 
exceeds the typical lifetime of protoplanetary disks.  
Instead, several alternative scenarios have been suggested. 
\citet{Pollack96-1} proposed that 
rocky planetesimals could grow to 
solid cores which are then scattered to larger separations 
where they accrete gaseous material to become the gas giant planets or brown
dwarfs at large separations we observe.  
Alternatively, SSCs could represent 
the low mass end of multiple stars
that formed similar to stellar binary stars, i.e. their formation mechanism
could be fragmentation of collapsing protostellar clouds
\citep[e.g.][]{Cha03-1}. 
Accretion in 
very young single brown dwarfs is consistent with direct fragmentation from
collapsing molecular cores \citep{thiesetal15-1}. SSCs 
could thus form the same way as single brown dwarfs with the only difference 
that they are members of multiple systems.

However, \citet{Kratter10-1}
pointed out that while this scenario offers a reasonable explanation for 
brown dwarf or stellar companions, it only works for the planetary mass 
companions if the companion forms at nearly exactly the
time the circumstellar envelope is exhausted which can be considered rather
unlikely.  
Another scenario that has been suggested for the formation of SSCs 
are gravitational instabilities (GI). 
If a massive gaseous disk becomes gravitationally unstable and
fragments into a  number of self-gravitating bound structures
\citep{Boss97-1}, these structures can then further collapse to become 
giant planets or brown dwarfs. However, models suggest that while the disk instability can 
indeed form planets at
separations 30-70 au \citep{Boss11-1}, it is unclear if the mechanism also works at
separations of several hundred au \citep[e.g.][]{rafikov07-1}.  
To overcome this problem, SSCs could form closer and be ejected to large
orbital separations \citep{whitworth+stamatellos06-1, stamatellosetal07-1}. 
Even clear predictions for objects formed by disk instabilities or an
isolated cloud have been presented 
\citep{stamatellos+herczeg15-1}. 
However, more recently \citet{mercer+stamatellos17-1} 
included radiation
feedback in the models and find that planetary mass companions
formed through disk instabilities are most likely 
ejected. 
In addition, \citet{Meru10-1,Kratter11-1}
find that gravitational instabilities only
occur for a certain and rather narrow set of 
conditions.

Despite the increasing number of SSCs detected at more 
than 100\,au from 
their host systems and masses below 30\,$M_\mathrm{Jup}$
\citep{Neuhauser05-1,Lafreniere08-1,Schmidt08-1,Deaconetal2016-1,Naudetal14-1},
we struggle to understand which formation mechanism is most appropriate. 
An important trait to study in order to progress with our understanding of SSC
formation is whether 
they are still accreting gas and at what rate this accretion proceeds. 
Accretion plays a key role in the formation of all stars, brown dwarfs, and
giant planets. 
Unfortunately, 
clear evidence based on spectroscopy for accretion in SSCs has been reported 
only for FW\,Tau\,b, CT\,Cha\,b and GSC\,06214-00210\,B 
\citep{Bowler14-1,Bowler11-1,Wu15-1}. 

The small number of accretion measurements for SSCs is caused by the fact 
that, despite their relatively large separations, classical spectroscopy of
SSCs is often impossible due to the large contrast at 
optical wavelength and limited spatial resolution 
in the infrared. 
One exception where spectroscopic 
accretion measurements are feasible 
is the brown dwarf
companion to the young stellar object SR\,12. 
SR\,12\,C is an SSC at a separation of 
more than 1000 au from the central binary.  
SR\,12\,AB is 
a binary T Tauri star 
which is located in the $\rho$ Ophiuchi star-forming 
region. 
The central binary, consists of stars with spectral type K4 and M2.5 
\citep{Bouvier+Appenzeller92-1,Gras05-1} with estimated masses of 1.05 
$M_{\sun}$ and 0.5 $M_{\sun}$. The separation of the two 
components is $\sim$0."21, equivalent to $\sim$26\,au.
The age of SR\,12 is unfortunately very uncertain. 
\citet[][]{Kuzuhara11-1} assumed that the age of
SR\,12 C corresponds to the age of young stellar objects in the low-extinction
region of $\rho$\,Oph and provided a rough estimate of
$2.1^{+7.9}_{-1.8}$\,Myr.  A younger age (0.016 Myrs) has been obtained by \citet{wahhajetal10-1} 
using the stellar evolutionary tracks from \citet{siessetal00-1} but these
authors themselves recommend to use the provided ages only 
to compare the relative ages of groups of objects.  
Throughout this work we thus assume an age of $\sim2$\,Myrs
keeping in mind that this value represents a very rough estimate.
The projected separation 
between SR\,12\,C and the central T Tauri star on the 
sky is $\sim$ 8.7", which corresponds to 
$\sim$ 1083 $\pm$ 217 au at 125 $\pm$ 25 pc \citep{geusetal89-1}. 
\citet{Kuzuhara11-1} obtained NIR spectra of SR\,12\,C with CISCO mounted on the 
Subaru telescope. They further showed that the 
probability of an arbitrarious alignment between SR\,12 C and SR\,12\,AB is 
$\sim$\,1 per cent. 
Based on theoretical age-luminosity relations the mass of SR\,12 C has been 
estimated to be 
0.013 $\pm$ 0.007 $M_{\sun}$ \citep[][]{Kuzuhara11-1}.  
\citet{Bowler14-1} observed SR \,12\,C 
with IRTF/SpeX and obtained spectra 
that matched an M9$\pm$ 0.5 spectral type, 
corresponding to a $T_{\mathrm{eff}} = 2400_{-100}^{+155}$K using 
$SpT-T_{\mathrm{eff}}$  relation provided by \citet{Luhman03-1}. 

SR\,12\,C is an ideal object to test for accretion 
in SSCs because of its large separation from the 
central binary. We here present X-shooter spectroscopy 
of SR\,12\,C and provide clear evidence for ongoing 
accretion based on several accretion indicators.
This makes SR\,12\,C just the fourth SSC with clear 
evidence for accretion. 
We also find that SSCs accrete 
at a rate similar to isolated objects 
and that accretion might perhaps cease at about 
the same age as in isolated
low-mass objects.



\section{Observations}
\label{sec:observations}
The large spectral coverage of X-Shooter \citep{vernetetal11-1}, extending from the UBV to NIR,
allows to probe simultaneously several accretion features
\citep[e.g.][]{Rigliaco12-1,manaraetal13-1,alcalaetal14-1}. We
observed SR12 with X-Shooter on the second of May 2016. 
The weather conditions were 
photometric with good seeing (less than 1"). 
We used slit widths of 1.3", 1.5" and 1.2" for the UVB, VIS and NIR
arm respectively. 
The exposure time was 4000\,s in total with a resolution of R$\sim$4000 for the UVB, 
R$\sim$5400 for the VIS, R$\sim$3890 for NIR. 
Although SR\,12\,C is separated from the central 
binary by a few times the seeing, the large flux contrast 
with the primary precludes us from detecting it on the 
acquisition images. To obtain the required spectrum we 
thus applied blind offsets to the acquisition 
images to place the slit on the companion using 
the position angle and separation values 
provided by \citet{Kuzuhara11-1}. 
The obtained data were reduced with the X-shooter pipeline using the
{\em{stare}} mode. 
The obtained SNR for the UVB were 2.3, 12.57 for the VIS arm, and 19.02 for the NIR arm.
Telluric correction and sky substraction was implemented with molecfit \citep{kauschetal15-1,smetteetal15-1} and
Skycorr \citep{nolletal14-1}. The $\sigma$ clipping method was implemented to remove pixels
deviating by more than 2.4 $\sigma$ from the median. 
To evaluate whether contamination from the central binary system 
affected our spectrum we fitted a combination of a Gaussian and a 
straight line to the 2D spectrum after binning 5 pixels in the dispersion 
direction. We do not find any evidence for a flux 
gradient in the background which clearly confirms that contamination from the
central binary is negligible.

\section{Characterizing SR\,12\,C}
\label{sec:char}
The obtained X-shooter spectrum of SR\,12\,C 
covers the UV to NIR simultaneously and 
allows us to derive tight constraints on the physical properties 
of this young brown dwarf. 
To that end we fit both theoretical and observational templates to the
X-shooter spectrum and derive spectral type, surface gravity, and temperature
of the SSC. 

\subsection{Fitting model templates}
\label{sec:fitting}
\begin{figure}
\includegraphics[width=\columnwidth]{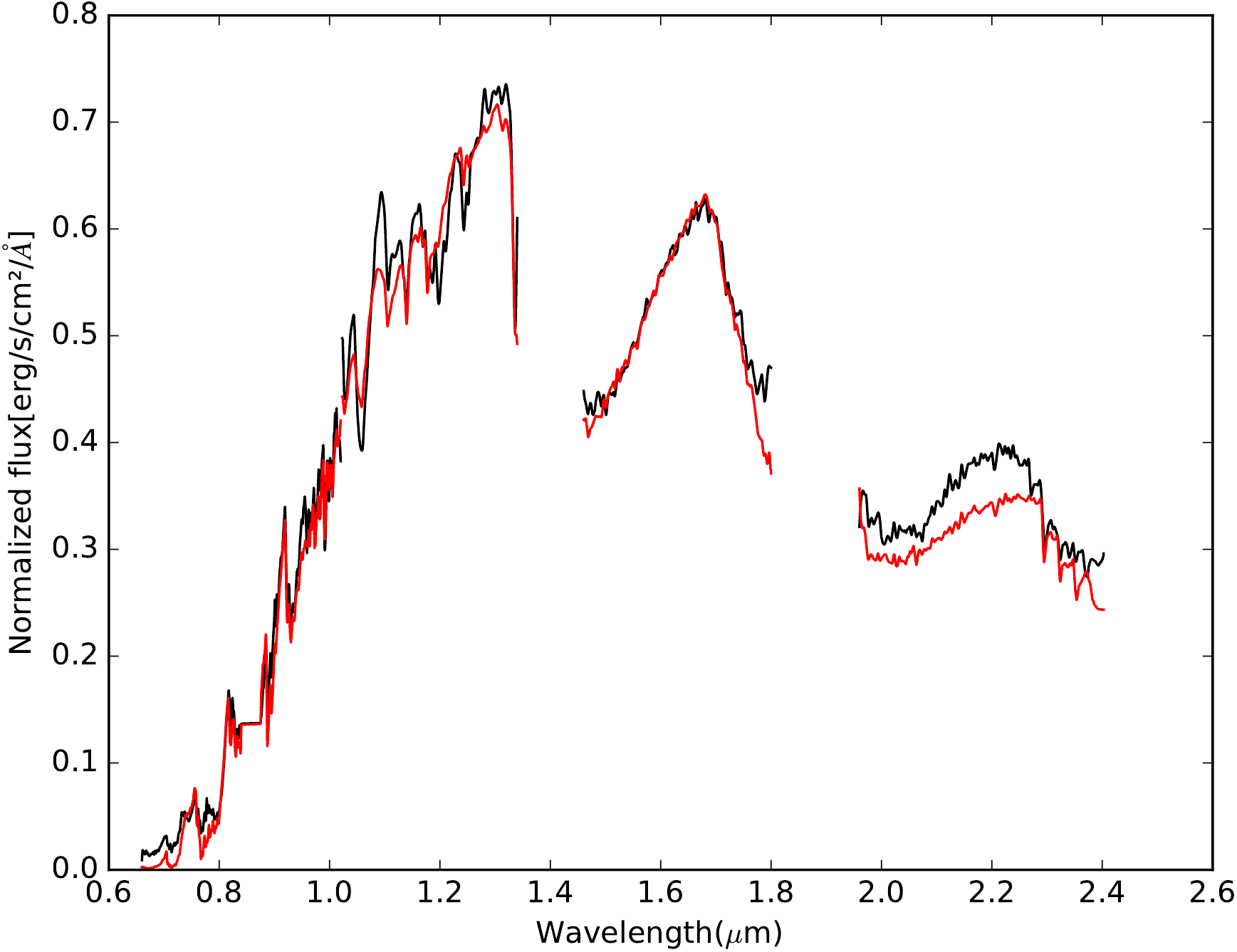}
  \caption{Theoretical spectrum (red line)
    fitted to the observed spectrum (black line). In general the agreement is
    reasonable with exception of the K band, 
where the influence of the dust creates a higher level of extinction.}
\label{fig:Baraffe}
\end{figure}

We used a grid of spectra based on 
BT-SETTL models \citep{allardetal14-1} 
to fit the X-shooter spectrum.  
These model spectra assume solar abundances 
and cover near infrared 
to visible wavelengths. 
Our grid covered values of $T_{\mathrm{eff}}$
ranging from 1600 to 3000 K and gravities 
between $\log$ g = 2.5 and 5.5
with a step sizes of 100\,K and $\log$\,g$=$0.5 respectively. 
To isolate the purely photospheric features that 
should be reproduced by the models,
we masked 
telluric regions 
in the IR and the most prominent 
emission lines, which were H$\alpha$, 
the Ca II triplet ($\lambda\lambda$8662, 8542, 8498 \AA), and the OI line 
at $\lambda$8446 \AA. 
We also resampled   
the theoretical spectra in order to 
obtain the same 
spectral resolution as the observed spectrum. 
We used the extinction law 
from \citet{Fitzpatrick99-1} with a fixed 
ratio
of total to selective 
extinction ($R_{V} = A_{V}/E_{B-V}=5.1$)
leaving $A_{V}$ as a free parameter to account for 
the intrinsic dust expected in the photosphere of 
brown dwarfs with effective temperature in the range from 2500 to 2700K.  
This dust and its effect on the emitted spectrum is known to be not accurately 
described by the theoretical models \citep{manjavacasetal14-1,bayoetal17-1}. 
We furthermore added a component for the 
interstellar extinction 
with $R_{V}=3.1$ and a fixed colour excess of $E_{B-V}=0.4$ 
that was obtained fitting only the visual part of the 
spectrum where the extinction curve is in the linear 
regime 
\citep[see Fig.\,1 in][]{Fitzpatrick99-1}.
For each theoretical spectrum we used least square
minimization with the extinction produced by the dust in the 
photosphere of the brown dwarf and the scaling factor as free 
parameters. 
The best fit had a temperature of 2600\,K 
and a surface gravity of  
$\log$ g= 4.0 with reduced $\chi^{2}=10.3$.
However, we note that models in the range of 
2500\,K to 2800\,K and with $\log$\,g between 3.5 and 4.5 
fit almost equally well the observations with 
$\chi^{2}$ values in the range of $\sim\,10.5-11.0$. 
Based on visual inspection focusing on the
shape of the spectrum in the J and H band,  
we finally estimate $\Teff$=2600$\pm$100\,K 
and $\log$ g= 4.0$\pm$0.5.
Combining these values for the 
effective temperature and the gravity with the isochrones from the BT-Settl 
models we derived an age estimate of 2 to 15 Myr (which is 
consistent with the 2 Myrs we assume throughout this work). The mass of
SR\,12\,C we estimate this way is slightly
larger (but still within 2$\sigma$) than previously estimated. 
Figure \ref{fig:Baraffe} compares the observed spectrum
and the best fit model. The observed spectrum is 
plotted in red while the best fitting model spectrum 
is represented by the black line. 
In the wavelength range from 0.66 to 2.0 $\mu$m the model reproduces well the observations. 
However, in the K band the model predicts too much 
flux. As mentioned above, 
this effect is well-known and generally explained by 
the presence 
of dust in the atmosphere of the brown dwarf which 
increases extinction and 
which is not properly included in the models
\citep{hiranakaetal16-1,manjavacasetal14-1,maroccoetal14-1}. 
Our simple addition of an extra-extinction can not 
entirely solve this issue
most likely because extinction laws assume grain size
distributions that do not need to match 
those of substellar atmospheres. 
In the H band the spectrum shows a triangular profile 
previously reported by \citet{Bowler14-1} and 
\citet{Kuzuhara11-1}.
This feature is a commonly 
known indicator of 
low surface gravity 
\citep{martin+osorio03-1,kirkpatricketal06-1}
in agreement with the age derived for SR\,12. 

\subsection{Observational template fitting}
We used two sets of observational templates to 
determine 
the spectral type of SR\,12\,C. 
First we used a library of near-infrared (1.1-2.45 \micron) integral field 
spectra of young M-L dwarfs with a resolution of 
R$\sim1500$ -- $2000$ \citep{Bonnefoy14-1}. These spectra were obtained with the 
Spectrograph for INtegral Field Observations in the Near Infrared (SINFONI) 
mounted at the VLT/UT4. 
The second set of spectra that we used was the SpeX prism Library
\citep{Rayneretal03-1} 
which is a repository of low-resolution spectra, primarily of low-temperature brown
dwarfs, obtained with the SpeX spectrograph on the 3m NASA Infrared Telescope
Facility on Mauna Kea, Hawaii. The wavelength coverage of these spectra is $0.65-2.55 \mu m$. 
We used both libraries as the SINFONI spectra have a resolution similar to our
observations but cover just the near-infrarred while the SpeX spectra  
cover the same wavelength range as our spectrum but with a significantly lower 
resolution. The combination of both libraries may provide the most accurate
constraints on the spectral type of SR\,12\,C. 

\begin{figure}
	\includegraphics[width=\columnwidth]{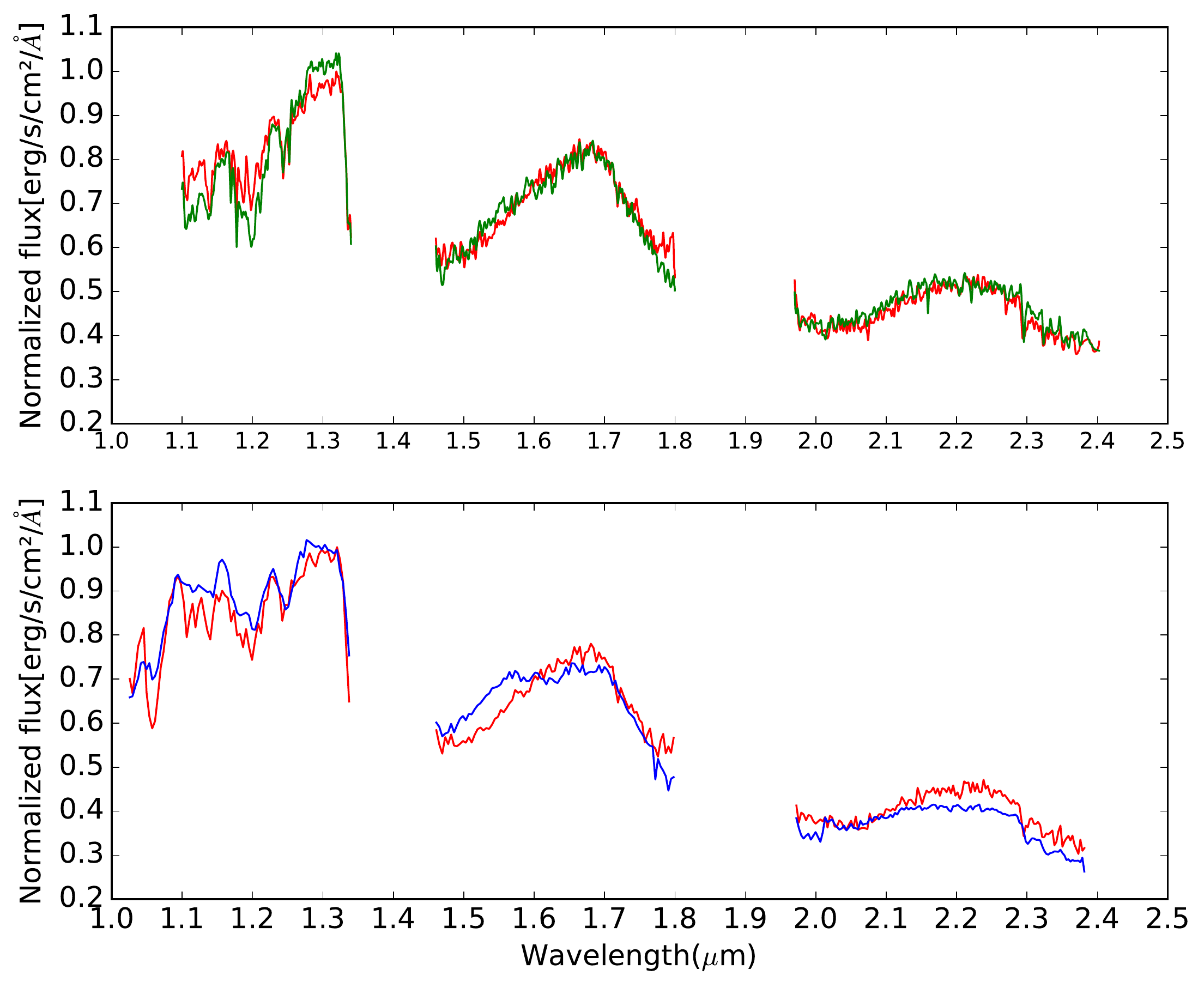}
    \caption{\textbf{Top panel:} Bonnefoy's L0 template(green line) versus SR\,12\,C(red line). \textbf{Bottom panel:} Spex Prism L0 template spectrum(blue line) versus SR\,12C(red line)}
    \label{fig:observacionales}
\end{figure}

In both cases we resampled the observed SR\,12\,C 
spectrum to obtain the same resolution as the templates. 
We also de-reddened SR\,12\,C with $A_{V} = 1.24$ and $R_{V} = 3.1$. The value for 
$A_{V}$ was obtained as explained 
in Sect. \ref{sec:fitting}.  
The H$\alpha$ line and the Ca II triplet were masked out
in the case of the SpeX prism Library. 
Using the SINFONI spectra we obtained the best fit 
($\chi^{2}$ = 5.6) with 2M40141, a brown dwarf 
classified as L0. 
While the spectrum of the M9.5 object OTS\,44   
still provides a reasonable fit ($\chi^{2}$ = 6.6) 
the fit gets much worse if we move to earlier spectral 
types ($\chi^{2}=14.8$ for 2M1207\,A with a spectral type M8.5). The library does not contain 
complete spectra for L1-L3 objects but the L4 object 
Gl417B does clearly not provide a good 
fit ($\chi^{2}$ = 9.4). 
The top panel of Fig.~\ref{fig:observacionales} 
compares the spectrum of 
SR\,12\,C (red line) with the scaled spectrum of 
2M40141 (green line).

The best fit ($\chi^{2}$ = 4.0) using the SpeX library 
was obtained with 
2MASP\,J0345432+2543023 
which is classified as an L0 
at optical wavelength and as an L1$\pm$1 brown dwarf 
in the NIR. 
While using M9 and L1 spectra available in the library 
(objects LHS 2924 and 2MASSW J1439284+192915) provides still reasonable
agreement ($\chi^{2}$ = 5.1 and 6.5 respectively), earlier or later spectral
types do clearly not provide acceptable approximations of our observations as
$\chi^{2}$ increases by more than a factor of three compared to our best fit.

The bottom panel 
of Fig.~\ref{fig:observacionales} shows the spectrum 
of SR\,12\,C 
(red line) versus a scaled version of 2MASP\,J0345432+2543023 (blue line).  
Both templates fit the K band much better than the 
theoretical spectrum which 
indicates that, despite significant recent progress, spectral models probably 
lack a proper description of the effects of dust in the atmosphere of brown
dwarfs \citep{manjavacasetal14-1}.  

Our results agree with previous studies. 
\citet{Kuzuhara11-1} and also \citet{Bowler14-1} obtained a $M9 \pm 0.5$ for
the spectral type of SR\,12\,C. 
Using the $\Teff$-spectral type relation 
provided by \citet{Luhman03-1}, our
spectral type determination converts to 
$2400^{+155}_{-100}$K, in reasonable agreement with the
value we found from fitting theoretical spectra. 
We conclude that SR\,12\,C should be classified 
as an L0$\pm$1 brown dwarf. 

\subsection{Radial velocity measurements}

We measured the radial velocity of SR\,12\,C using two different
  features, the calcium triplet emission lines ($\lambda\lambda$8662, 8542,
  8498 \AA) and the sodium absorption doublet ($\lambda\lambda$5889.9, 5895.9
  \AA). We fitted the triplet and the doublet simultaneously 
and found radial velocities of   
-6.1 $\pm$  1.0 km/s (calcium triplet) -7.4 $\pm$ 2.8 km/s (sodium doublet). 
We also reduced an archival spectrum of the central binary (SR\,12\,AB) which 
had been taken in the framework of ESO project 093.C-0506(A). 
This spectrum was taken with a position angle of 161.462 degrees and a 
total integration time of 120 seconds. We fitted the same lines for the 
host star and we obtained the following radial velocities: 
-6.7 $\pm$  0.5 km/s (calcium triplet) and -7.1 $\pm$  0.5 km/s (sodium
doublet). We conclude that there is no radial velocity difference 
between SR\,12\,C and the central binary system. This indicates that the 
formation scenario in which PMCs form close to the central object and are 
then ejected does probably not apply to SR\,12\,C. 

\subsection{Identification of absorption lines}

\begin{figure*}
	\includegraphics[width=18cm]{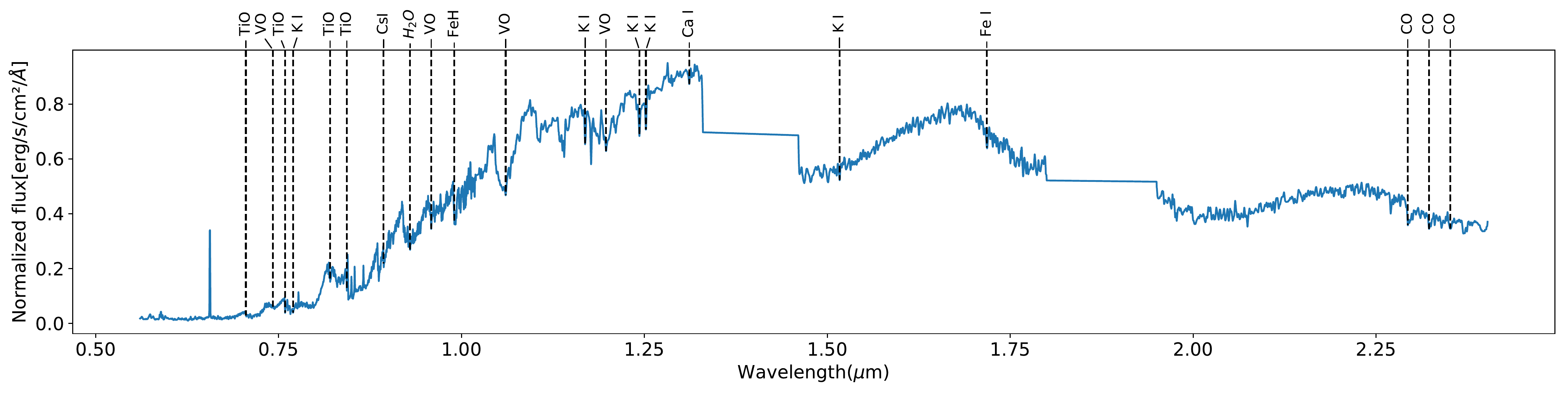}
    \caption{Spectral identification of the most common features in late M and early L types.}
    \label{fig:features}
\end{figure*}

While the main focus of this work is on accretion signatures and possible
constraints on formation theories of PMCs, for the sake of completeness 
we investigated which photospheric absorption lines are present in our 
X-shooter spectrum. We used Fig.\,2 of \citet{bayoetal17-1} as a reference 
for typical spectral features in late M and L dwarfs 
\citep[see
  also][]{jonesetal94-1,kirkpatricketal93-1,geballeetal96-1,allardetal97-1}. We
identified several in the spectrum of SR\,12\,C that are common in 
late M and early L stars (see Fig. \ref{fig:features}). 
A complete list of these lines together with 
the measured equivalent widths is given in the appendix.



\section{Accretion} 
\label{sec:acc}
As outlined in the introduction, an important 
parameter for constraining 
formation scenarios for SSCs is the age range in which 
these objects accrete and the associated mass 
accretion rates. 

The X-shooter spectrum of the brown dwarf 
orbiting SR\,12 contains clear evidence for ongoing accretion. 
According to the magnetospheric accretion model \citep{Koning91-1} the stellar 
magnetosphere truncates the disk near the surface of the star. Material from
the inner disk is transferred through magnetic field lines as accretion 
columns onto the high latitude regions of the star where it dissipates its 
kinetic energy in standing shocks \citep{Calvet98-1}. Shock fronts at the base 
of the accretion columns create a high temperature layer 
\citep{Koning91-1,Gullbring94-1,Lamzin95-1} of $\sim 10^{4}$K where the 
optically thick post shock gas and optically thin pre-shock gas generate 
emission leading to broad 
velocity profiles of $H\alpha$ and the CaII IR triplet.  
Other emission lines that are produced in the 
shocks at the base of the accretion columns and that are frequently used to
measure accretion are the Paschen and the Balmer series, and Oxygen lines.
We identified all these accretion tracers in the X-shooter spectrum
of SR\,12\,C and measured 
the accretion rates using eight of them.

\subsection{Accretion estimated from $H\alpha$ emission}
H$\alpha$ is the most commonly used accretion indicator 
in Classical T-Tauri stars(CTTS). 
Accretion rates seem to correlate well with the H$\alpha$ 10 per cent width 
\citep{Natta04-1}
where 
values $\gappr$ 200 km/s \citet{Jayawardhana03-1} or 
$\gappr$ 270 km/s \citep{ciezaetal10-1}
are considered as evidence for accretion \citep[see
  also][for a discussion]{romeroetal12-1,ciezaetal12-1}. 
Less broad emission lines can be produced by chromospheric 
activity and cannot be interpreted as evidence for
accretion.  

\begin{figure}
	\includegraphics[width=\columnwidth]{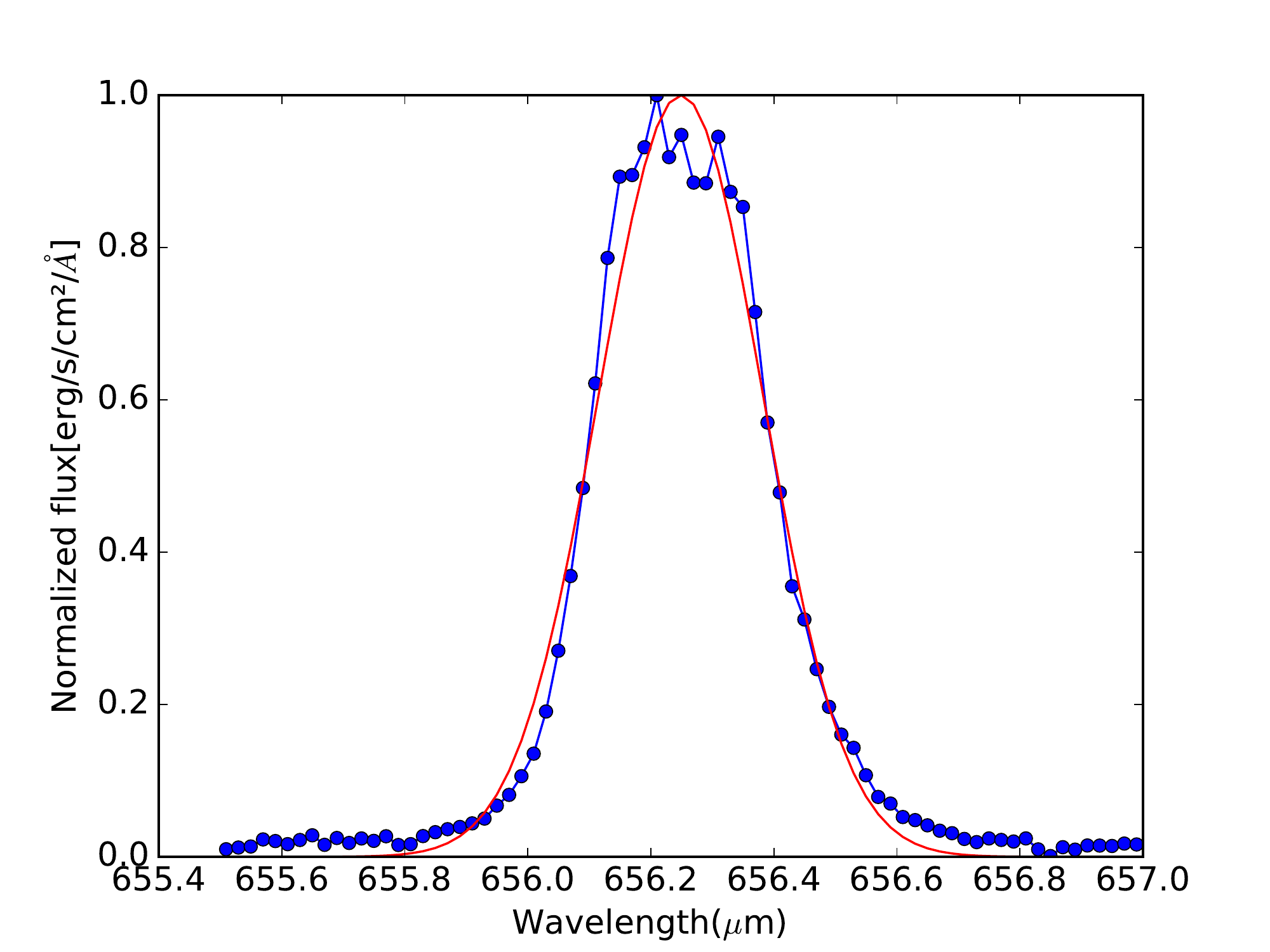}
    \caption{The observed H$\alpha$ line (blue)
      is fitted with a Gaussian profile (red). The irregular shape and top
      part of the line is related to magnetospheric accretion. The width at 10 per cent of the line corresponds to
an accretion rate of $\dot{M}_{\mathrm{acc}} = 10^{-10.31 \pm 0.5} M_{\sun}/\mathrm{yr}$}
    \label{fig:halfa}
\end{figure}

Although this diagnostic was originally developed for CTTS 
it can be used for objects below the hydrogen burning limit as well. 
In brown dwarfs, H$\alpha$ has been used to measure accretion 
adopting the limit of $\sim$ 200 km/s and using the 
following equation
\begin{equation}
 \log \dot{M}_{\mathrm{acc}} = -12.89(\pm 0.3) +9.7(\pm0.7) \times 10^{-3} H\alpha 10 \%
 \label{eq:macc1}
\end{equation}
where $H\alpha$ is in km/s and $\dot{M}_{\mathrm{acc}}$ in $M_{\sun}/yr^{-1}$
\citep{Natta04-1}. 
Using this method we obtain 266.3 km/s as the 10 per cent width which correspond to a
$\dot{M}_{\mathrm{acc}} = 10^{-10.31 \pm 0.5} M_{\sun}/yr$ (see Fig. \ref{fig:halfa}). 
\label{sec:alpha} 

\begin{table*}
	\centering
	\caption{Characteristics of the observed accretion tracers.}
	\label{tab:intermediate_results}
	\begin{tabular}{lcccc} 
		\hline
		Line (\AA) & EW (\AA) & Flux line (W/$m^{2}$) & $\log$ ($L_{line}$/ $L_{\sun})$ & $L_{\mathrm{acc}}$		($L_{\sun}$)\\
		\hline
		H$\alpha$ $\lambda$ 6562.8  & $-57.39 \pm 17.71$ &$(7.69\pm 2.39)\times 10^{-17}$ & $-4.43 \pm 2.65$ & $(2.24 \pm 0.98 )\times 10^{-4}$ \\
		H$\beta$ $\lambda$ 4861  & $-46.90 \pm 4.21$ &$(2.19 \pm 0.20 )\times 10^{-17}$ & $-4.97 \pm 2.15$ & $(2.22 \pm 0.64 )\times 10^{-4}$\\
		H$\gamma$ $\lambda$ 4341  & $-23.06 \pm 2.28$ & $(6.57 \pm 0.66 )\times 10^{-18}$ & $-5.50 \pm 2.16$ & $(1.50 \pm 0.42 )\times 10^{-4}$\\					
		H11 $\lambda$ 3771  & $-9.37 \pm 0.44$ & $(8.93 \pm 0.43 )\times 10^{-19}$ & $-6.36 \pm 2.11$ &$(6.26 \pm 1.66 )\times 10^{-5}$\\
		OI $\lambda$ 8446  & $-6.97 \pm 0.50$ & $(1.12 \pm 0.90 )\times 10^{-18}$ & $-6.27 \pm 2.14$ &$(6.09 \pm 1.47 )\times 10^{-4}$\\
		\hline
	\end{tabular}
\end{table*}

\subsection{Accretion measured with the Ca\,II triplet}
\begin{figure*}
	\includegraphics[width=16cm]{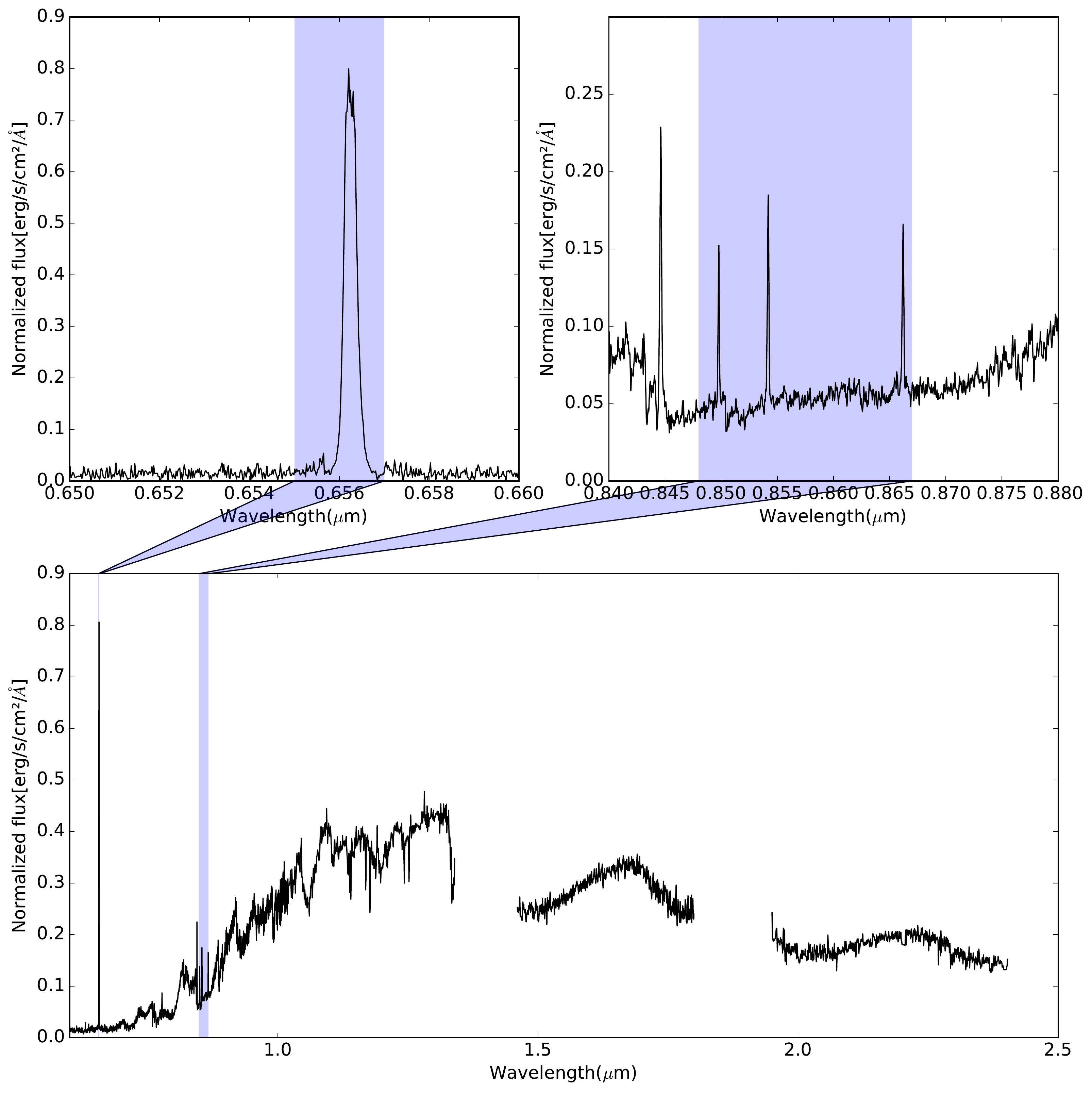}
    \caption{The X-shooter spectrum (covering the visible and 
J, H, K bands) of SR\,12 C shows clear evidence for 
ongoing accretion (bottom pannel). In the
top panels we highlight the strong H$\alpha$ emission
      line (left) and CaII triplet and OI line (right). 
}
    \label{fig:anel}
\end{figure*}

The CaII triplet emission lines are another frequently used accretion
indicator in young stellar objects. These lines are present 
in SR\,12\,C and the line flux ratio clearly indicates accretion 
(see Fig.\,\ref{fig:anel}). 
As shown by \citet{Comeron03-1}, if the line flux ratio is close to 1:1:1, 
the lines cannot be
produced by chromospheric activity because optically thin emission would cause line flux
ratios of 1:9:5. Measuring the line flux ratio of the CaII
triplet is therefore one of the most reliable accretion tracers. 
We measured the line flux ratio and obtained 1:1.37:1.09, a
clear sign of ongoing accretion in SR\,12\,C. 

Furthermore, \citet{Comeron03-1} derived simple equations relating 
the accretion rate to the line flux based on 
the theoretical work of \citet{Muzerolle98-1}. These relations are: 
\begin{equation}
\log \dot{M}_{\mathrm{acc}} = -34.15 + 0.89 \times \log(F_{CaII(\lambda 8542)})
 \label{eq:macc2}
\end{equation} 
\begin{equation}
\log F_{CaII(\lambda 8542)} = 4.72 \times 10^{33} EW($\AA$) \times 10^{-0.4(m_{\lambda}-0.54A_{V})} 
 \label{eq:macc3}
\end{equation} 
where $F_{CaII(\lambda8542)}$ is the line flux, $m_{\lambda}$ is the 
magnitude of the star at $\lambda$8542, and $A_{V}$ is the visual extincton 
at the wavelength of the line. 
Using the above relations we obtained an accretion rate 
of $4.83 \times 10^{-10}$ $M_{\sun}yr^{-1}$ 
for SR\,12\,C. 


\begin{figure*}
	\includegraphics[width=16cm]{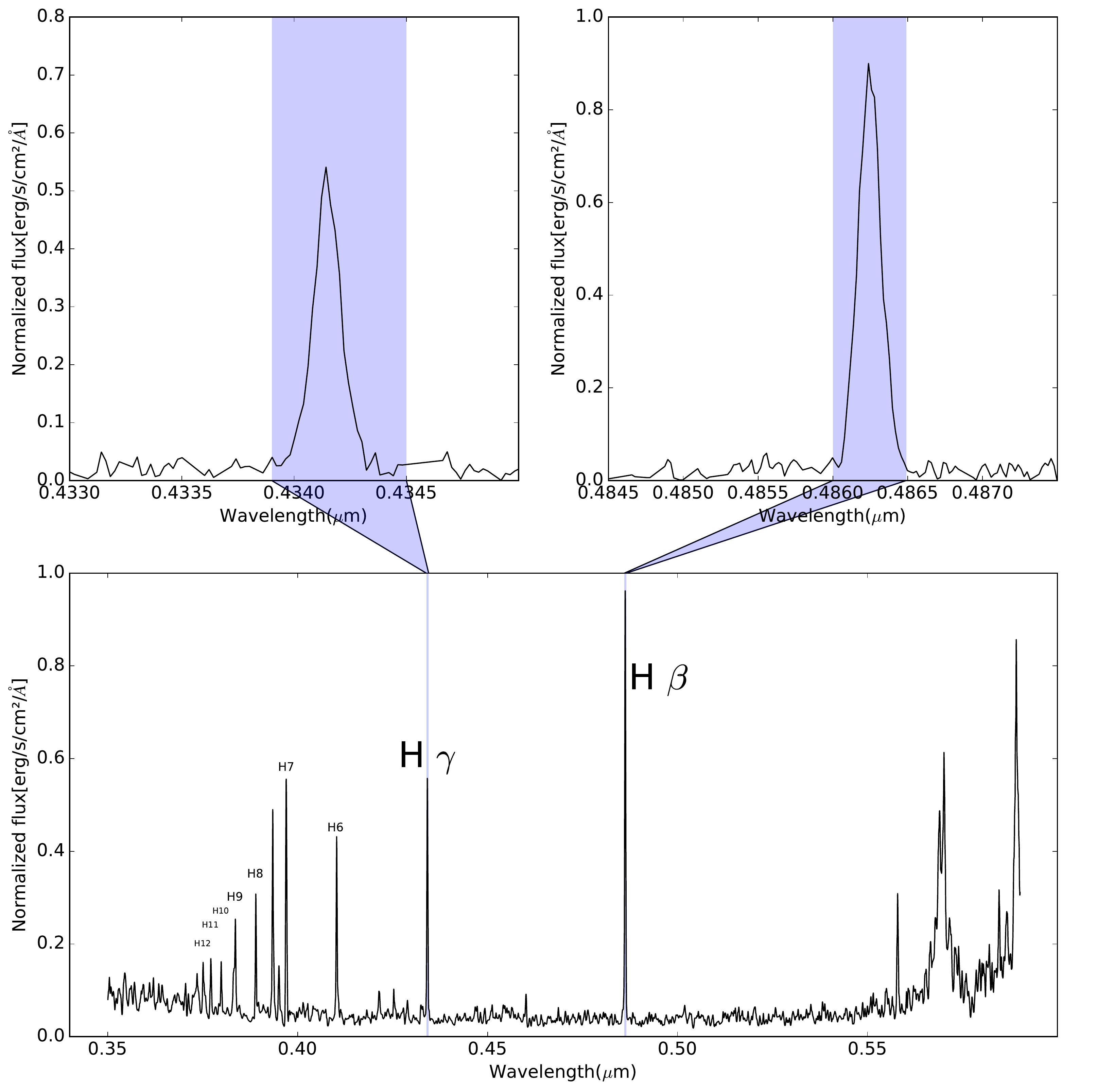}
    \caption{Bottom pannel: SR\,12\,C uvb
      spectra, with the Balmer series labeled. Top left pannel: H $\gamma$ emission
      line. Top right pannel: H$\beta$ emission line. }
    \label{fig:balmer}
\end{figure*}
\begin{table}
	\centering
	\caption{Accretion rates measurements for each emission line}
	\label{tab:rates}
	\begin{tabular}{lcc} 
		\hline
		Line & Wavelength (\AA) & $\dot{M}_{\mathrm{acc}}$ ($M_{\sun} yr^{-1}$)\\
		\hline
		H$\alpha$ & $6562.8^{a}$ & $(4.90 \pm 5.64 )\times 10^{-11}$\\
		H$\alpha$ & $6562.8^{b}$ & $(1.16 \pm 0.88 )\times 10^{-11}$\\
		H$\beta$ & 4861 & $(1.04 \pm 0.74 )\times 10^{-10}$\\
		H$\gamma$ & 4341 & $(7.01 \pm 4.98 )\times 10^{-11}$\\					
		H11 & 3771 & $(2.93 \pm 2.06 )\times 10^{-11}$\\
		OI & 8446  & $(1.02 \pm 0.75)\times 10^{-10}$\\
		CaII & 8544.2 & $4.83\times 10^{-10}$\\
		Pa$\beta$ & 12800 & $(3.95 \pm 2.94 )\times 10^{-13}$\\
		\hline
	\end{tabular}
    \begin{tablenotes}
      \small
      \item \textbf{Notes} \\
      $^{a}$ Measured as explained in Section ~\ref{sec:alpha} \\
      $^{b}$ Measured as explained in Section ~\ref{sec:Hlines}
    \end{tablenotes}
\end{table}

\begin{table*}
	\centering
	\caption{Wide (>100\,au) companions under 15 $M_{Jup}$ and younger than 10 Myrs}
	\label{tab:pmc_accretion}
	\begin{tabular}{lcccccccc} 
		\hline
		Object & Mass  & Age & Separation & SpT & Primary  & Accretion$^{b}$ & Accretion rate & References \\
		& ($M_{Jup}$) & (Myr) & (AU) & & $\mathrm{Multiplicity}^{a}$ & & log$\dot{M}$($M_{\sun} \mathrm{yr}^{-1}$)\\
		\hline
		1RXS J1609-2105b & $8^{+4}_{-2}$ & $\sim$5 & $\sim$330 & L2$\pm$1 & S &	N & - & 1,2,3,4,5\\
		ROXs 42B b & 9$\pm$3 & $7^{+3}_{-2}$ & $\sim$140  & L1$\pm$1 & B & U & - & 6,7\\
		FW Tau b & 6-14 & $2^{+1}_{-0.5}$ & $\sim$330 & - & B & Y & -11.0$\pm 1.3^{c}$ & 6,7,8\\
		ROXs 12 b & 12-20 & $8^{+4}_{-3}$ & $\sim$210 &-& S & $N^{d}$ & - & 7,9\\
		CHXR 73 B & $13^{+8}_{-6}$ & $\sim$2 & $\sim$210 & $\geqslant$ M9.5 & S & U & - & 10,11\\
		DH Tau B & $12^{+10}_{-4}$ & $\sim$1-2 & $\sim$330 & M9.25 $\pm$ 0.25 & S & Y & -9 to -11 & 9,12,13,14,30\\
		GSC 6214-210 B & $14\pm2$ & $\sim$5 & $\sim$320 & M9.5 $\pm$ 1 & S & Y & -10.7$\pm$ 1.3 & 5,6,15 \\
		CT Cha B  & 17$\pm$6 & 0.9-3 & 440 & - & S & Y & -$\sim$9.22 & 16,32\\
		WD 0806-661B  & 8$\pm$ 2 & 1.5$\pm$0.3 & 2500 & Y & S & U & - & 17,18\\
		HD 106906 b  & 11$\pm$2 & 13$\pm$2 & 654$\pm$3 & L2.5 $\pm$ 1 & S & U & - & 19, 2\\
		Ross 458 C  & 11$\pm$4.5 & 150-800 & 1168 & T8.5$\pm$0.5 & B & U & - & 20,21,22,23,24,25 \\
		AB Pic B  & 13$\pm$0.5 & 0.9-3 & 275 & L0 $\pm$ 1 & S & U & - & 26,27\\
		SR 12 C & 13$\pm$7 & $\sim$2 & 1083$\pm$217 & L0 $\pm$ 1& B & Y & -9.84$\pm$0.53 & 6,28,29 \\
				Oph 1622-2405 & 21$\pm$3 & 11$\pm$2 & 243$\pm$55 & $\geqslant$ L8 & S & U & - &2,33 \\
		 	UScoCTIO 108 b & $16^{+3}_{-2}$ & 11$\pm$2 & $\sim$670 & M9.5& S & U & - & 2,34 \\
		\hline
	\end{tabular}
	  \begin{tablenotes}
      \small
      \item \textbf{Notes} \\
      Table based on Table\,1 from \citet{Bowler14-1}
      $^{a}$ Single("S") or Binary ("B")
      $^{b}$ Yes("Yes"), No("N"), Unknown or non reported ("U")
      $^{c}$ \citet{Bowler14-1} They also calculated log $\dot{M}$= 11.4 $\pm$
      1.3 ($M_{\sun} yr^{-1}$) depending on the calculation of the Mass.
      $^{d}$ \citet{Kraus14-1} describes H$\alpha$ line weak as a WTTS, accretion was discarded.
      GSC 6214-210 B and FW TAU B accretion rate are based on Paschen $\beta$.
      In the case of DH TAU B the accretion rate is based on its optical
      excess emission. Accretion rate of CT Cha B measured from excess
      emission in r'.
      \item \textbf{References.} \\
      (1) \citet{Wu15-1};(2) \citet{Pecaut12-1};(3) \citet{Lafreniere08-1};(4) \citet{Lafreniereetal10-1};(5) \citet{Ireland11-1};(6) \citet{Bowler14-1};(7) \citet{Kraus14-1};(8)\citet{Caceres15-1};(9)\citet{Bouvier+Appenzeller92-1};(10)\citet{Luhman04-1};(11)\citet{Luhman06-1};(12)\citet{Itoh05-1};(13)\citet{White+Ghez01-1};(14)\citet{Bonnefoy14-1};(15)\citet{Bowler11-1};(16)\citet{schmidtetal08-1}; (17) \citet{rodriguezetal11-1}; (18) \citet{luhmanetal11-1}: (19) \citet{Bailey14-1}; (20) \citet{goldmanetal10-1}; (21) \citet{dupuyetal13-1}; (22)\citet{scholzetal12-1}; (23) \citet{burgasser10-1}; (24) \citet{burninghametal11-1};(25) \citet{beuzitetal04-1}; (26)\citet{chauvinetal05-1}; (27)\citet{perrymanetal97-1}; (28)\citet{Kuzuhara11-1};(29) this work; (30) \citet{zhouetal14-1}; (31) \citet{Neuhauser05-1}; (32) \citet{Wu15-2}; (33) \citet{closeetal07-1}; \citet{bejaretal07-1}
     \end{tablenotes}
\end{table*}

\subsection{Accretion rate based on Paschen $\beta$}
\label{sec:paschen_section}
One of the regularly used accretion tracers in low-mass stars is 
Paschen $\beta$.  
\citet{Natta04-1} found that the Pa$\beta$ luminosity correlates with the 
accretion luminosity. 
We measured the Pa$\beta$ 
luminosity in SR\,12\,C 
and obtained  
$\log(L_{Pa\beta}/L_{\sun}) = -7.16 \pm  2.37 \Lsun$. The empirical relation 
found by \citet{Natta04-1} and revised by \citet{Rigliaco12-1}
is $\log L_{\mathrm{acc}} = (1.49\pm 0.04 )\times \log L(Pa\beta)+ 4.59 \pm 0.14$ which gives 
$L_{\mathrm{acc}} =(-8.46 \pm 3.00) \times10^{-7}$ $\Lsun$ for SR\,12\,C.  
Assuming a mass of SR\,12\,C of $0.013 \pm 0.007 \Msun$  
\citep{Kuzuhara11-1}
and a radius of $0.19\pm0.07 \Rsun$ (derived from the gravity values we
obtained by spectral fitting) finally results in a mass accretion rate of 
$\dot{M} =(3.96\pm 2.94)\times10^{-13}  M_{\sun}$yr$^{-1}$.

\subsection{Accretion rate based on the OI line}
Additional emission lines that can be related to accretion are 
those from OI \citep[e.g.][]{joergensetal12-1}. 
Our X-shooter spectrum contains a strong OI $\lambda$8446 
emission line that we here use to add another measurement of the 
accretion rate in SR\,12\,C. 
First, we calculate the line luminosity in a similar fashion as in 
the previous subsection. 
We then use Fig.\,10 from \citet{Herczeg08-1} to estimate the accretion
luminosity of SR\,12\,C (note that the coefficients for the shown linear
correlation are not given) and finally obtain an accretion rate of 
  $\dot{M} =(2.85\pm 1.98)\times10^{-10}  M_{\sun} yr^{-1}$ using the 
mass and radius of SR\,12\,C as above. The given uncertainty does not 
take into account the standard deviation of the linear fit from 
\citet{Herczeg08-1} as these values are not available to us. However, the included
uncertainties related to the mass and radius of SR\,12\,C likely dominate
the error estimate. 

\subsection{Accretion rates derived from other Hydrogen lines}
\label{sec:Hlines}
The entire Balmer and Paschen emission line series 
has been used to measure accretion \citep{Ferguson97-1}. 
We used $\log (L_{\mathrm{acc}}/L_{\sun}) = b + a \times \log (L_\mathrm{line}/L_{\sun})$ to
relate the accretion luminosity and the luminosities of the lines 
(see \citealt{Fang09-1,Herczeg08-1} and Table 6 and 8 from
\citealt{Rigliaco12-1}). 
We applied this method 
to the H$\alpha$, H$\beta$, H$\gamma$ and H11 lines identified in the spectrum
of SR\,12\,C. 
Figure \ref{fig:balmer} shows the UV spectrum (bottom panel) of SR\,12\,C. 
Equivalent width, flux line (W/$m^{2}$), $\log$($L_{Pa \beta}$/ $L_{\sun})$ and
accretion luminosity in solar units are given 
in Table \ref{tab:intermediate_results}. 
The resulting accretion rates can be found in Table \ref{tab:rates}. 

\begin{figure}
	\includegraphics[width=\columnwidth]{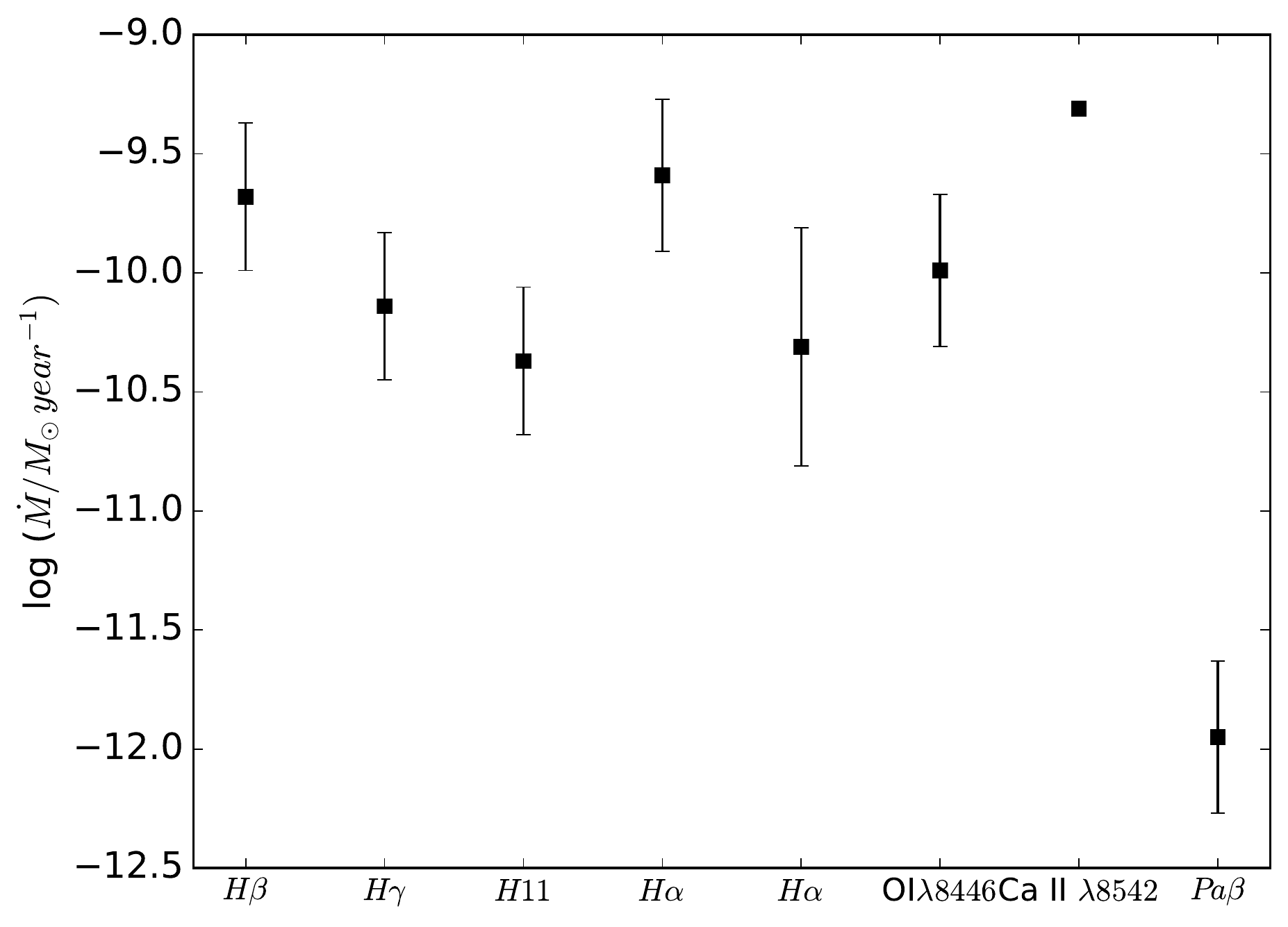}
	\caption{Accretion rates obtained from different emission
      lines. Apparently, the obtained values agree very with each other. The
only exception is the accretion derived from Pa$\beta$ which is significantly
      lower. 
	}
	\label{fig:accretion_8}
\end{figure}
\begin{figure}
	\includegraphics[width=\columnwidth]{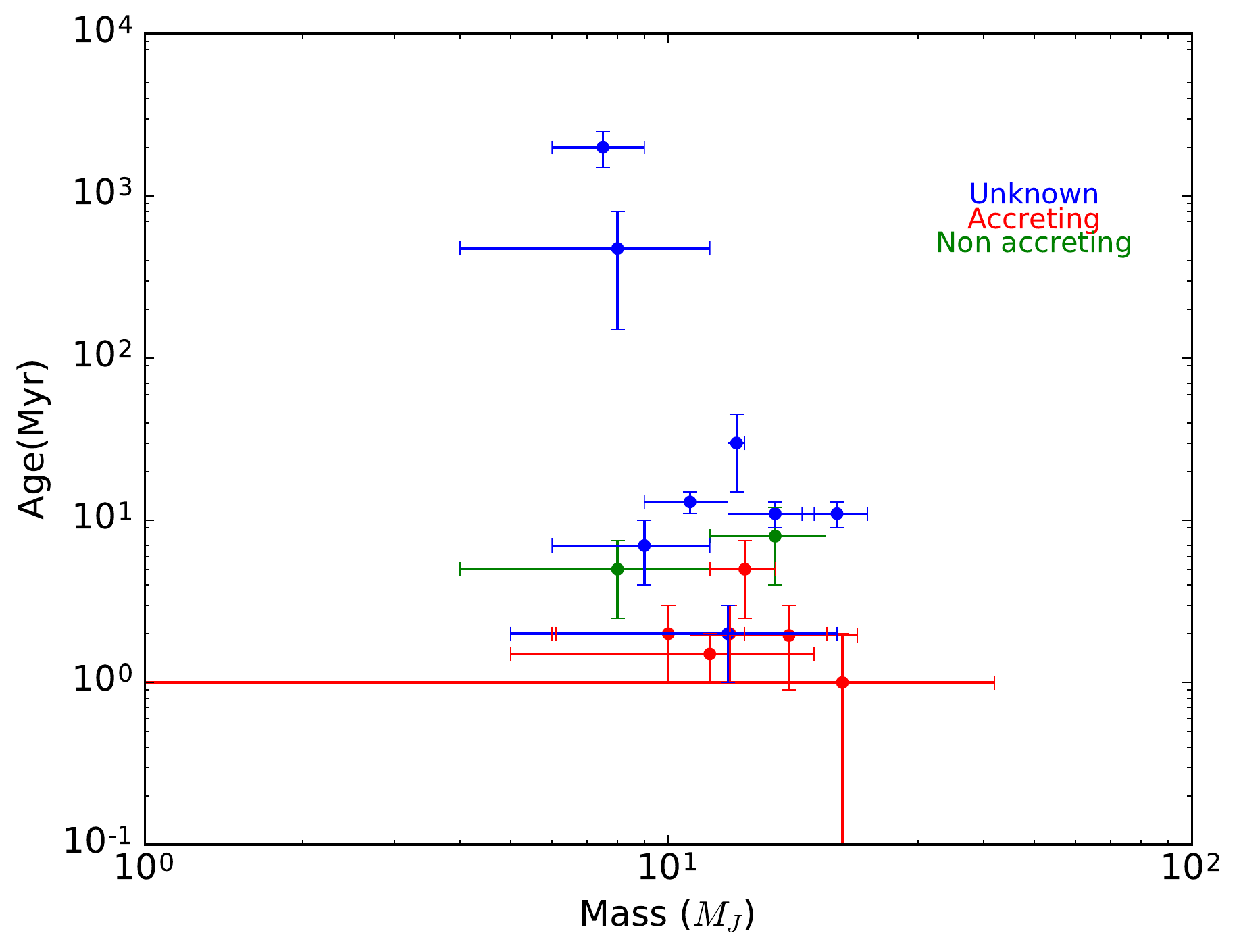}
	\caption{Age in Myrs versus mass of the SSCs listed in \ref{tab:rates}.
Red indicates that the objects are accreting, green represents non accreting 
objects and in blue we show SSCs without information on their accretion 
status. It seems that objects younger than 10 Myrs are usually accreting. }
	\label{fig:masa_age}
\end{figure} 
\begin{figure}
	\includegraphics[width=\columnwidth]{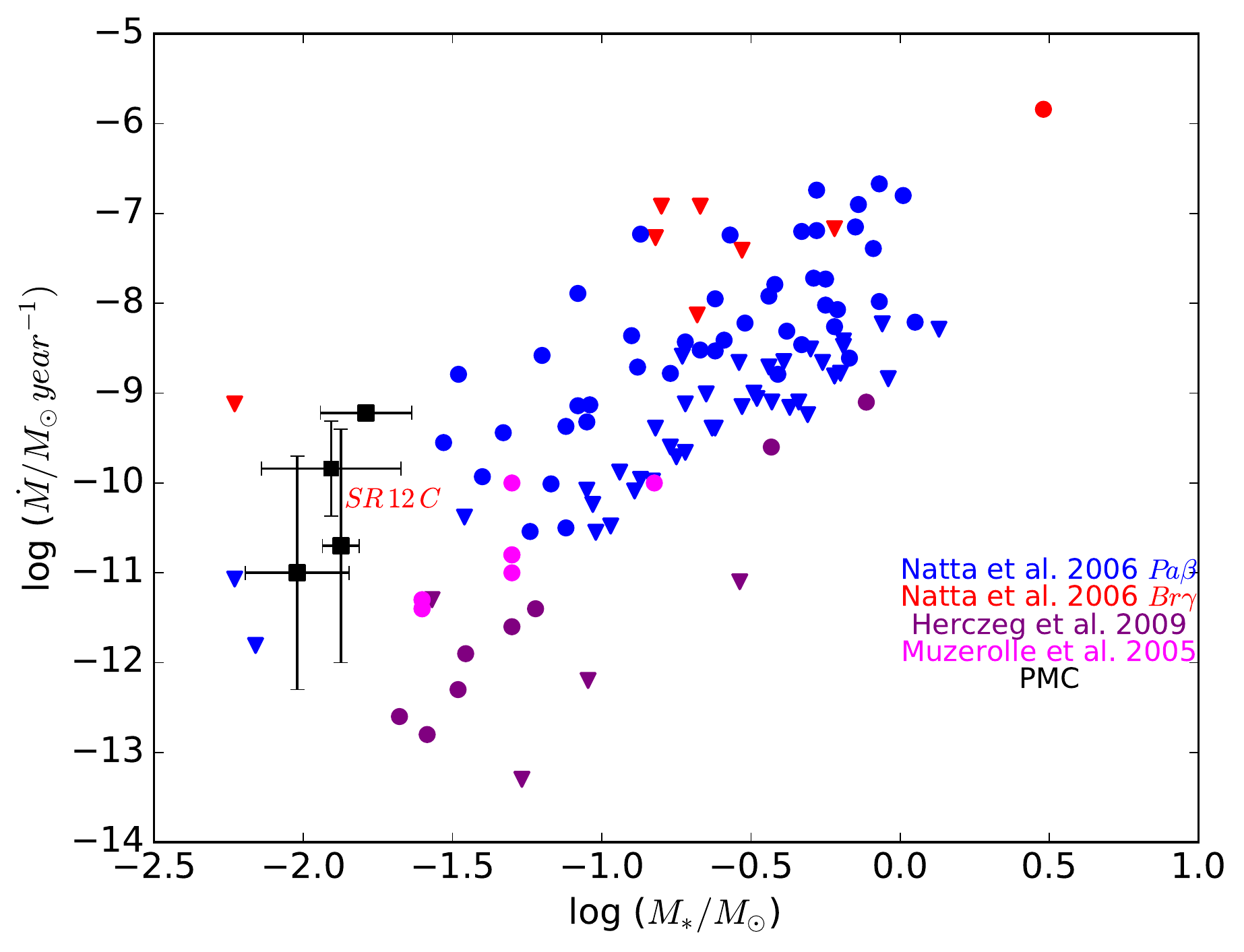}
	\caption{Based on Fig.\,6 from \citet{Bowler11-1}. Mass accretion rates vs
      stellar mass. Triangles shows upper limits. We include FW TAU b, GSC
      06214-00210 b, CT Cha B and SR\,12\,C as black squares. Data shown in
      blue are accretion rates given in \citet{natta+randich06-1} which 
are measured using $Pa\beta$ while those in red are measured with
$Br\gamma$. Purple symbols represent measurements from \citet{herczegetal09-1}
and green pentagons have been obtained from \citet{muzerolleetal05-1}. We
selected only class II objects from \citet{natta+randich06-1}. Keep in mind
that this plot combines accretion rate measurements with different
methodologies of objects with different ages. However, it seem that SSCs
roughly follow the correlation between mass and accretion rate of young stellar objects.}
	\label{fig:masa_macc}
\end{figure}

\section{Discussion}
\label{sec:discussion}

We have measured accretion rates for the substellar companion (SSC) to the
young binary SR\,12 using eight different methods. 
In general, the measured accretion rates are very similar and agree with each
other. However, the accretion rate measurement based on Pa $\beta$ is
significantly lower than the other values we obtain, as illustrated in 
Fig.\,\ref{fig:accretion_8}. As all values agree very well except the one
obtained from Pa$\beta$, the empirical relation used for
Pa$\beta$ seems to provide accretion rates that are perhaps less
    reliable. 
This hypothesis is somewhat supported by looking at the accretion rates 
obtained for other SSCs. Accretion rate measurements exist for 
three other SSCs(CT\,Cha\,b,GSC 6214-210\,B and FW Tau\,B). For CT\,Cha\,b
\citet{Wu15-1} estimate a relatively high accretion rate of
$\log\dot{M}$($M_{\sun} yr^{-1}$)=-9.22 from interpreting 
excess emission in the r'\,-band as caused by H$\alpha$ emission. 
This is similar to the value we obtained for SR\,12\,C using all techniques
(i.e. $\log\dot{M}(M_{\sun} yr^{-1})\sim-10$) except the one based on Pa$\beta$.  
For GSC 6214-210\,B and FW Tau\,B published accretion rate measurements 
are based on Pa$\beta$ emission and the obtained values are 
relatively low, i.e. $-10.7 \pm 1.3$ and $-11.0 \pm 1.3 $ respectively. 
Thus one might get the impression that indeed be that Pa$\beta$ accretion rates of SSCs are 
systematically lower. 
However, \citet{joergensenetal13-1} investigated accretion on the isolated
brown dwarf OTS\,44 and find that the rate derived from Pa$\beta$ is
significantly {\em{larger}} than the one obtained from H$\alpha$. 
Thus, further simultaneous measurements of accretion rates based on
different lines are needed 
before any firm conclusions can 
be drawn. Nevertheless, we conclude that the higher values we obtain 
for SR\,12\,C  with 8 different methods probably represent a more reliable 
estimate of the accretion rate in SR\,12\,C. 
The currently available accretion rate measurements for SSCs
as listed in Table \ref{tab:pmc_accretion}. 
In addition to the three objects mentioned above, 
two more objects have accretion rate estimates based on the accretion 
luminosity from excess continuum emission \citep{zhouetal14-1}.  

In Fig.\,\ref{fig:masa_age} we show accreting and non accreting
systems as a function of age and mass. 
Objects in red are those that are accreting, objects
in green have been reported as non-accreting in the literature, and the
objects in blue lack accretion measurements. 
While more measurements are clearly needed before we can draw firm
conclusions, this figure illustrates that SSCs younger than 10 Myr seem to be 
generally accreting. A similar tendency is observed in isolated brown
dwarfs. For example, 
\citet{liuetal03-1} identify a decreasing disk fraction 
around low-mass objects and brown dwarfs supporting the idea that
disks do usually not survive longer than 10\,Myr. 
The potential similarity between isolated low-mass objects and 
SSCs is further supported by
Figure \ref{fig:masa_macc} where we show the position of the SSCs with 
accretion measurements in the $\log\dot{M}-\log M$ diagram together with  
young isolated low mass stars and brown dwarfs. SSCs seems to follow the 
usual correlation between mass an accretion rate \citep[see also][]{Bowler14-1}. 

In summary, SSCs seem to share two accretion traits with young isolated brown
dwarfs. These are their 
similar accretion rates,
as well as the observation that both are accreting if they are younger than 
10 Myrs. Although we are in an early stage concerning the understanding of these 
objects and only few observational constraints are available, the measured
accretion rates might indicate that SSCs perhaps form 
in a similar way as isolated brown dwarfs. 
For example, if SSCs are originally formed close to the host star 
due to core accretion, one would expect the circumplanetary disks to be
severely affected \citep[see][for details]{Bowler11-1} and would thus not
expect the accretion on the companions to follow the same trend as observed in
single low-mass objects. For the remaining two 
scenarios, 
disk instabilities and collapsing protostellar clouds,  
the predicted accretion rates are quite similar 
\citep{stamatellos+herczeg15-1}. 
The currently available measurements accretion rates 
of SSCs do not allow to distinguish between the two 
scenarios although the larger accretion rates predicted by disk instability
models would perhaps even fit slightly better.  
However, if radiation feedback is included 
disk fragmentation has general problems 
in producing planetary mass companions at large 
separations \citep{mercer+stamatellos17-1}. 
The least problematic formation scenario therefore seems to be 
collapsing prestellar clouds, i.e. 
SSCs represent just 
the low-mass end of companions in multiple stellar systems.

\section{Conclusion}
\label{sec:conclusion}


SR\,12\,C is a substellar companion orbiting the binary system SR\,12\,AB at 
a projected separation of 1083 $\pm$ 217 au in the $\rho$ Ophiuchus star forming cloud. 
We observed SR\,12\,C using X-Shooter
and fitted the resulting spectrum with a set of theoretical templates and 
two different
sets of observational templates and find that SR\,12\,C is best described as
brown dwarf with surface gravity $\log g = 4$, a temperature of 2600\,K 
and that its spectral type is L0. These results agree well with previous
studies of the object. 

Investigating several accretion indicators we find 
that our X-shooter spectrum provides clear evidence for ongoing accretion in
SR\,12\,C. We estimated the accretion rate using different emission lines and
different methods and find that the obtained values cluster around 
$\sim\,10^{-10}\Msun/\mathrm{yr}$.

Comparing the accretion rate measured for SR\,12\,C with those 
obtained for other SSCs and  
young isolated brown dwarfs we do not find indications for any
significant differences. Indeed, it seems that most SSCs 
accrete if they are younger than 
$\sim10$\,Myrs and with similar rates as young isolated low-mass objects. While further
observations of SSCs are clearly needed, this might perhaps indicate that the
formation mechanism for isolated brown dwarfs and low-mass stars also 
produces substellar companions.

\section*{Acknowledgements}
This research has benefitted from the SpeX Prism Spectral Libraries,
maintained by Adam Burgasser. 
ASM, CC, MRS thank for support from  the Millennium Science Initiative 
(Chilean Ministry of Economy), through grant RC13007. 
MRS also achnowledges support from Fondecyt (1141269). 
CC acknowledges support from CONICYT PAI/Concurso nacional de insercion en la academia 2015, Folio 79150049.
A. Bayo acknowledges financial support from the Proyecto Fondecyt Iniciaci\'on 11140572



\bibliographystyle{mnras}

\bibliography{bibliografia} 

\appendix

\section{Equivalent width of photospheric features in SR\,12\,C}

We identified the several absorption features common in late M and early L
stars. The lines and the equivalent width are given in the table below. 

\begin{table}
    \centering
    \label{tab:equivalent_width_features}
    \begin{tabular}{lcc} 
        \hline
        Line & Wavelength (\AA) & Equivalent Width (\AA) \\
        \hline
        TiO& 7053 & 0.83  $\pm$ 0.04\\
        VO & 7424 & 0.14 $\pm$ 0.01 \\
        TiO & 7589 & 0.59 $\pm$ 0.01 \\
        K I & 7697 & 0.47 $\pm$ 0.07\\                   
        TiO & 8204 & 0.62 $\pm$ 0.01\\
        TiO & 8433  & 0.72 $\pm$ 0.03\\
        FeH & 9901 & 0.82 $\pm$ 0.27\\
        VO & 9586 & 0.21 $\pm$ 0.01 \\
        CsI & 8935 & 1.33 $\pm$ 0.09 \\
        VO& 10604  & 0.31$\pm$ 0.01\\   
        K I& 11690  & 1.41 $\pm$ 0.16\\
        K I & 12432 &1.26 $\pm$ 0.02\\
        VO & 11975 & 0.37 $\pm$ 0.01\\
        K I & 12520 & 1.78 $\pm$ 0.05\\                   
        Ca I & 13113  & 2.08 $\pm$ 0.01\\
        K I & 15167  & 1.81 $\pm$ 0.13\\
        Fe I & 17181 & 0.28 $\pm$ 0.01\\
        CO & 22934 & 2.10 $\pm$ 0.06\\
        CO & 23225 & 1.02 $\pm$ 0.06\\
        CO & 23516 & 0.86 $\pm$ 0.06 \\

        \hline
    \end{tabular}
\end{table}


\end{document}